\documentclass[twocolumn,pre,floatfix]{revtex4}

\RequirePackage{color}
\usepackage{color}
\usepackage{epsfig}
\usepackage{latexsym,  mathrsfs}
\usepackage{amsmath, amssymb}
\usepackage{makeidx, url}
\usepackage{enumerate}
\newtheorem{algo}{Algorithm}[section]
\newcommand{\tline}{\rule{\linewidth}{0.1mm}}
\newcommand{\til}[1]{\tilde #1}
\newcommand{\norm}[1]{\| #1 \|}

\begin{document}

\title{Parallel  Self-Consistent-Field Calculations 
       via Chebyshev-Filtered Subspace Acceleration
}

\author{
Yunkai Zhou$^{(a)}$\footnote{corresponding author, yzhou@smu.edu},
Yousef Saad$^{(b)}$,
Murilo L. Tiago$^{(c)}$,
and James R. Chelikowsky$^{(c,d)}$
}

\affiliation{
$^{(a)}$Department of Mathematics,
Southern Methodist University,
Dallas, TX 75275,  USA. \\
$^{(b)}$Department of Computer Science \& Engineering, 
 University of Minnesota,
 Minneapolis, MN 55455, USA. \\
$^{(c)}$ Center for Computational Materials, Institute
  for Computational Engineering and Sciences, University of Texas,
  Austin, TX 78712, USA. \\
$^{(d)}$ Departments of Physics and Chemical Engineering,
  University of Texas, Austin, TX 78712, USA.
}

\date{\today}

\begin{abstract}
Solving  the   Kohn-Sham  eigenvalue  problem   constitutes  the  most
computationally  expensive   part  in  self-consistent   {\em  density
functional theory} (DFT) calculations. In a previous paper,
we have proposed a nonlinear 
Chebyshev-filtered subspace  iteration  method, which  avoids  
computing  explicit
eigenvectors except  at the first  SCF iteration.  The method  may be
viewed  as  an approach  to  solve  the  original nonlinear  Kohn-Sham
equation  by   a  nonlinear  subspace   iteration  technique,  without
emphasizing the intermediate linearized Kohn-Sham eigenvalue problems.
It reaches self-consistency  within a  similar number  of SCF
iterations  as eigensolver-based  approaches.  However,  replacing the
standard diagonalization at each SCF iteration by a Chebyshev subspace
filtering step results in a  significant speedup over methods based on
standard   dagonalization.    Here, we discuss an approach for implementing this method in multi-processor, parallel environment.
Numerical  results are
presented to show that the method enables to perform a class of highly
challenging DFT calculations  that were not feasible before.
\end{abstract}

\maketitle

\section{Introduction} \label{introd}

Electronic structure calculations based on first principles
use a very successful  combination of 
{\it density functional theory} (DFT) \cite{hkohn:64,kosh:65}
and {\it pseudopotential theory} 
\cite{phillips-58,phil-klein-59,chelik-cohen-92,martin_book_04}.
DFT reduces the original multi-electron Shr\"{o}dinger equation
into an effective one-electron Kohn-Sham equation, where all 
non-classical electronic interactions are replaced by a functional
of the charge density. The pseudopotential theory 
further simplifies the
problem by replacing the true atomic potential with an effective 
``pseudopotential'' that is smoother but takes into account the
effect of core electrons.
Combining pseudopotential with DFT greatly reduces the number of
one-electron wave-functions to be computed.
However, even with these simplifications, solving the final Kohn-Sham 
equation can still be computationally challenging, especially
when the systems being studied are complex or contain thousands of
atoms.

Several approaches have been employed in solving the Kohn-Sham
equations. They can be
classified in two major groups: basis-free or basis-dependent
approaches, according to whether they use an explicit basis set for
electronic orbitals or not. Among the basis-dependent approaches, plane-wave
methods are frequently used in applications of DFT to periodic systems
\cite{payne:92,vasp96}, whereas localized basis sets are very popular
in quantum-chemistry applications \cite{martin_book_04,koch_book_00}. 
Special basis sets, which do not make use of pseudopotentials, 
have also been designed for all-electron DFT calculations. These basis
sets include localized atomic orbitals,
linearized augmented plane waves, muffin-tin orbitals, and
projector-augmented waves. A survey of advantages and disadvantages
of these explicit-basis methods can be found in
\cite{martin_book_04}. Real-space methods are basis-free,
and they
have gained ground in recent years 
\cite{cts:94,ctws:94,real-space-cg,beck-00} due in great part
to their simplicity. One advantage of real-space methods
is that they are quite easy to implement in parallel
environment. A second advantage is that,  in contrast with
the plane-wave approach, they do not 
impose artificial periodicity in non-periodic systems.
Third, the application of potentials onto electron wave-functions is
performed directly in real space. 
Although the Hamiltonian matrices with a real-space
approach are typically
larger than with plane waves, the Hamiltonians are highly
sparse and never stored or computed explicitly. 
Only matrix-vector products that represent the application of the 
Hamiltonians on wave-functions need to be computed.

This article
focusses on  effective techniques to  handle  the most
computationally  expensive  part   of  DFT  calculations,  namely  the
self-consistent-field   (SCF)  iteration.  
We  present details of
a  recently developed  nonlinear  Chebyshev-filtered  subspace iteration  
(CheFSI) method.   
The sequential version of CheFSI is first proposed in \cite{chefsi}.
The parallel CheFSI is
implemented   in  our   own   DFT   package  called   PARSEC
(Pseudopotential Algorithm for Real-Space Electronic Calculations)
\cite{cts:94,ctws:94}. 
Although CheFSI is described in the framework of real-space DFT, 
the  subspace filtering method can be employed to other Self-Consistent 
Field iterations. 
This  method takes advantage of the fact that intermediate SCF iterations 
do not require accurate eigenvalues and eigenvectors of the Kohn-Sham
equation.

The  Standard  SCF  iteration  framework  is used  in  CheFSI,  and  a
self-consistent solution  is sought, which  means that CheFSI  has the
same accuracy as other standard DFT approaches.
One can  view CheFSI  as a technique  to directly tackle  the original
nonlinear  Kohn-Sham  eigenvalue  problems  by  a  form  of  nonlinear
subspace iteration,   without emphasizing the  intermediate linearized
Kohn-Sham  eigenvalue  problems.  In  fact,  within  CheFSI,  explicit
eigenvectors are  computed only at the first  SCF iteration,
in order to provide a  suitable initial subspace.  After the first SCF
step, the  explicit computation of eigenvectors at  each SCF iteration
is replaced by  a single subspace filtering step.   The method reaches
self-consistency within  a number of  SCF iterations that is  close to
that of eigenvector-based approaches.  However, since eigenvectors are
not explicitly  computed after the  first step, a significant  gain in
execution time  results when compared  with methods based  on explicit
diagonalization. 
When compared with calculations based on 
efficient eigenvalue packages such as ARPACK \cite{lesy:98} 
and TRLan \cite{trlan99,wusi:00},  a tenfold or higher  speed-up 
is usually observed. 
CheFSI enabled us to perform  a class of  highly challenging DFT
calculations, including  clusters with over ten  thousand atoms, which
were not feasible  before.  

This article begins with a summary of SCF for DFT calculations in
Section \ref{eigp}. Details about the parallel implementation are included
in Section \ref{ppp}. The Chebyshev subspace filtering algorithm is
presented in Section \ref{main-alg}, and the block Chebyshev-Davidson
algorithm for the initial diagonalization is discussed in Section \ref{chdav}. 
The block Chebyshev-Davidson method \cite{chebydav,bdavio} improves considerably
the efficiency of the  diagonalization at the first SCF iteration, compared 
with the thick-restart Lanczos (TRLan) method \cite{trlan99,wusi:00}
which was used in \cite{chefsi}.  
The paper ends with numerical results in Section \ref{numer}, and a few
concluding remarks.

\section{Eigenvalue problems in DFT SCF calculations} \label{eigp}
Within DFT, the multi-electron Schr\"{o}dinger equation is
simplified as the following Kohn-Sham equation:
\begin{equation}\label{kseq}
\left[ - \frac{\hbar^2}{2M} \nabla^2 + V_{total}(\rho(r), r) 
\right]     \Psi_i(r) = E_i \Psi_i(r),
\end{equation}
where $\Psi_i(r)$ is a wave function, $E_i$ is a Kohn-Sham
eigenvalue, $\hbar$ is the Planck constant, and $M$ is the electron mass. 
In practice we use atomic units, thus $\hbar = M = 1$.

The \emph{total potential} $V_{total}$, also referred to as 
the {\it effective potential}, includes three terms,
\begin{equation}\label{totalv}
 V_{total}(\rho(r), r) = V_{ion}(r) + V_H(\rho(r), r) +
 V_{XC}(\rho(r), r),
\end{equation} 
where $V_{ion}$ is the ionic potential,
$V_H$ is the Hartree potential, 
and $V_{XC}$ is the exchange-correlation potential.

The Hartree and exchange-correlation potentials depend 
on the {\it charge density} $\rho(r)$, which is defined as
\begin{equation}\label{density}
\rho(r) = 2 \sum_{i=1}^{n_{occ}} |\Psi_i(r)|^2.
\end{equation} 
Here $n_{occ}$ is  the number of occupied states, which is equal to half 
the number of valence electrons in the system. 
The factor of two comes from spin multiplicity.  
Equation  (\ref{density})  can  be easily  extended  to
situations where the highest occupied states have fractional occupancy
or when there is an imbalance in the number of electrons for each spin
component.

The most computationally expensive step of DFT is in solving the 
Kohn-Sham equation \ref{kseq}. Since  $V_{total}$  depends on the charge
density $\rho(r)$, which in turn depends on the wavefunctions $\Psi_i$,
this equation can be viewed as a \emph{nonlinear eigenvalue problem}.
The SCF iteration is a general technique used
to solve this nonlinear eigenvalue problem.
It starts with an initial  guess  of the  charge
density,  then  obtains the initial $V_{total}$  and
solves  \ref{kseq}  for   $\Psi_i(r)$'s  to   update   $\rho(r)$  and
$V_{total}$. Then \ref{kseq} is  solved again for the new $\Psi_i(r)$'s
and  the  process is  iterated  until $V_{total}$
(and also the wave functions) becomes stationary.   

In general, most of the computational effort involved in DFT is spent
solving equation \ref{kseq}. For this reason, it is the goal of any DFT code to
lessen the burden of  solving  \ref{kseq} in the SCF iteration. 
One possible avenue to achieve this is to use better
diagonalization routines. However this approach is limited
as most diagonalization software has now reached  maturation.
At the other extreme, one can attempt to avoid diagonalization
altogether, and this leads to the body of work represented by 
linear-scaling or order-N methods
(see e.g. \cite{goed99}). This approach however has other limitations.
In particular, the approximations involved rely heavily on 
some decay properties of the density matrix in certain function
bases. In particular, they will be difficult to implement in real-space
discretizations. Our approach lies somewhere between these extremes.
We take advantage of the fact that accurate eigenvectors are 
unnecessary at each SCF iteration, since Hamiltonians are only
approximate in the intermediate SCF steps,  
and we exploit the nonlinear  nature of the problem. 
The main point of our algorithm, developed in \cite{chefsi},  
is that once we have a good starting
point for the Hamiltonian, it suffices to filter each basis vector
at each iteration. 
In the intermediate SCF steps,
these vectors are no longer eigenvectors but together
they represent a good basis of the desired invariant subspace. 
The parallel implementation of the idea will be discussed in 
Section \ref{main-alg}. The next section summarizes  parallel 
implementation issues in PARSEC.

\section{The parallel environment in PARSEC} \label{ppp}
PARSEC 
uses pseudopotential real-space implementation of DFT.
The motivation and original ideas behind the method go back to the 
early 1990s \cite{cts:94,ctws:94}.
Within PARSEC, an uniform Cartesian grid in real-space is placed on the 
region of interest, and the Kohn-Sham equation is discretized 
by a high order finite-difference method \cite{fd-acta94} on this grid.
Wavefunctions are expressed as functions of grid positions.
Outside a specified sphere boundary that encloses the physical system,
wavefunctions are set to zero for non-periodic systems. 
In addition to the advantages mentioned in the introduction,
another advantage of the real-space approach is that periodic boundary
conditions are also reasonably simple to implement \cite{wjkc:04}.

The latest version of PARSEC is written in Fortran 95.
PARSEC has now evolved into a mature, massively parallel package,
which includes most of the functionality of comparable
DFT codes \cite{parsec-k}.
The reader is referred to  \cite{Saad-al-Elec,sosck:00} for details
and  the rationale of the parallel implementation.
The following is a brief summary of the most important points.

\begin{figure}[htpb]
{\centering\epsfig{figure=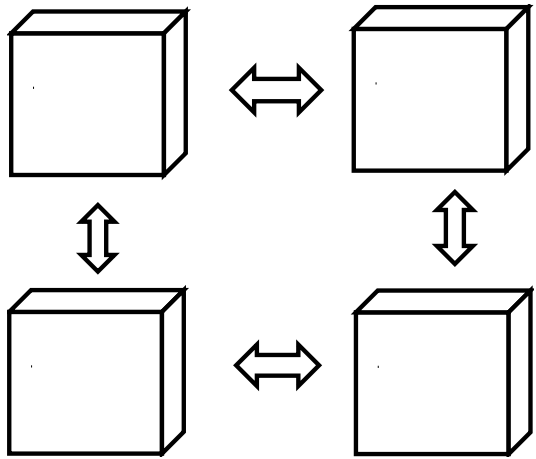,width=7cm,clip=}}
\caption{Sample decomposition of a physical domain in PARSEC. \hfill}
\label{ddm}
\end{figure}

The parallel mode of PARSEC uses the standard 
Message Passing Interface (MPI)  library for communication. 
Parallelization is achieved by  partitioning the physical domain which
can have various shapes  depending on boundary conditions and symmetry
operations.  Figure \ref{ddm} illustrates four cube-shaped neighboring
sub-domains.   For a  generic, confined  system without  symmetry, the
physical  domain  is a  sphere  which  contains  all atoms  plus  some
additional  space  (due to  delocalization  of  electron charge).   In
recent years, PARSEC  has been enhanced to take  advantage of physical
symmetry.   If   the  system   is  invariant  upon   certain  symmetry
operations, the physical domain  is replaced with an irreducible wedge
constructed according to those  operations. For example, if the system
has mirror  symmetry on the  $xy$ plane, the irreducible  wedge covers
only  one hemisphere,  either above  or  below the  mirror plane.  For
periodic  systems, the  physical domain  is the  periodic cell,  or an
irreducible wedge  of it  if symmetry operations  are present.  In any
circumstance, the  physical domain is partitioned  in compact regions,
each assigned to  one processor only. Good load  balance is ensured by
enforcing that the compact  regions have approximately the same number
of grid points.

Once the physical domain is partitioned, the physical problem is mapped
onto the processors in a data-parallel way: each processor is in
charge of a block of rows of the Hamiltonian corresponding to the
block of grid points assigned to it. The eigenvector and potential vector
arrays are row-wise distributed in the same fashion. The program only requires 
an index function $indx(i,j,k)$ which returns the number of the processor 
in which the grid point $(i,j,k)$ resides.

Because the Hamiltonian matrix is never stored, we need an explicit 
reordering scheme which renumbers rows consecutively from one
processor to the next one. For this purpose we use a list of pointers 
that gives for each processor, the row with which it starts. 

Since   finite difference  discretizetion is used, when 
performing an operation such as a matrix-vector product, 
communication will be required between nearest neighbor processors.
For communication we use two index arrays,
one to count how many and which rows are needed from 
neighbors, the other to count the 
number of local rows needed by neighbors.

With this design of decomposition and mapping, the data required by
the program can be completely distributed. Being able to distribute the memory 
requirement is quite important in solving large problems on standard 
supercomputers.

Parallelizing subspace methods for the linearized eigenvalue problems
(obtained from a finite difference discretization of 
eqn. \ref{kseq}) becomes quite straightforward with the 
above mentioned decomposition and mapping. Note that the subspace basis 
vectors contain approximations to eigenvectors, therefore the rows of the
basis vectors are distributed in the same way as the rows of the Hamiltonian.
All matrix-matrix products, matrix-vector products,
and vector updates (e.g., linear combinations of vectors), 
can be executed in parallel.

Reduction operations, e.g., computing inner
products and making the result available in each processor,
are efficiently handled by the 
MPI reduction function {\tt MPI\_ALLREDUCE()}.

\section{The nonlinear Chebyshev-filtered subspace iteration} \label{main-alg}

The main idea of CheFSI is to start with a good initial
subspace $V$ corresponding  to  occupied states  of  the  initial
Hamiltonian. This initial $V$ is usually obtained by a diagonalization step.
No diagonalizations are necessary after the first SCF step.
Instead, the subspace from the previous iteration is filtered by a
low degree-$m$ Chebyshev polynomial, $p_m(t)$, constructed  for the current
Hamiltonian $H$.  The polynomial differs at each
SCF step since $H$ changes. 
The goal of the filter is 
to make the subspace spanned  by  $p_m(H) V$  approximate the  
eigensubspace corresponding to the occupied states of the final $H$.
At the intermediate SCF steps, the basis  need not be an accurate
eigenbasis  since the intermediate Hamiltonians are not exact.
The  filtering  is designed  so  that the resulting sequence of subspaces  
will  progressively  approximate  the desired
eigensubspace  of  the  final  Hamiltonian  when  self-consistency  is
reached.  

Our approach exploits the well-known fast growth property outside the $[-1,1]$ 
interval of the Chebyshev polynomial, this allows us to use 
low degree Chebyshev polynomials to achieve sufficient filtering.
At each SCF step, only two parameters  are required to construct  
an effective Chebyshev filter, namely,
a lower bound and an upper bound of the higher portion 
of the spectrum of the  current Hamiltonian $H$  in which we
want $p_m(t)$  to be small. We propose simple but efficient
ways to obtain these bounds, very little
additional cost is required for the bound estimates.

After self-consistency is reached, the Chebyshev filtered 
subspace   includes  the   eigensubspace  corresponding   to  occupied
states.  Explicit  eigenvectors  can  be  readily obtained  by  a  {\it
Rayleigh-Ritz  refinement} 
\cite{parlet:98} (also called {\it subspace rotation}) step.

We refer to \cite{chefsi,tr-pchefsi} for more algorithmic details and a 
literature survey concerning application of Chebyshev polynomials 
in DFT calculations.

The   main  structure  of the CheFSI method is  given   in  Algorithm
\ref{chefsi}. It is quite  similar to that of the  standard SCF iteration
discussed in Section II.  One major difference is  that the inner
iteration for diagonalization at {\it Step 2} is now performed only at
the  first SCF  step.  Thereafter, diagonalization  is  replaced by  a
single Chebyshev-Filtered Subspace step, denoted as {\bf CheFS} in 
Algorithm \ref{chefsi}. 

\begin{figure}[pht]
\begin{algo}  CheFSI for SCF calculation: 
  \label{chefsi} \\
\tline 
\vspace*{0.1cm}
\begin{minipage}[r]{\columnwidth}
{\baselineskip=5mm
\begin{enumerate}[~~1.~]
\item Start from an initial guess of $\rho(r)$, get ~$V_{total}(\rho(r),r)$.

\item Solve ~$ 
\left[ 
 - \frac{1}{2}{\nabla^2} + V_{total}(\rho(r), r) 
\right]
      \Psi_i(r) = E_i \Psi_i(r) $ 
      ~ for $\Psi_i(r), \; i = 1, 2, ..., s$. 

\item Compute new charge density ~$\rho(r) = 2 \sum_{i=1}^{n_{occ}} |\Psi_i(r)|^2$.

\item Solve for new Hartree potential ~$V_H$ from 
      ~~$ \nabla^2 V_H(r) = -4 \pi \rho(r)$.

\item Update ~$V_{XC}$; get new
  ~${\til V}_{total}(\rho,r) = V_{ion}(r) + V_H(\rho, r) +
  V_{XC}(\rho, r)$ with \\ 
  a potential-mixing step.

\item If ~$\norm{{\til V}_{total} - V_{total}} < tol$, {\bf stop}.

\item  $V_{total} \leftarrow {\til V}_{total}$  (update $H$ implicitly);  
 apply the following \\
 Chebyshev-Filtered Subspace ({\bf CheFS}) method 
 to get $s$ approximate wavefunctions: 
\begin{enumerate}[7.1)~]
\item Compute  ~$b_{up} := $ upper bound of the spectrum of $H$ \\
 Set $b_{low} := $ the largest Ritz value from previous iteration.

\item 
Perform Chebyshev filtering to the matrix $  \Psi $, whose column-vectors\\
are the  $s$  discretized wavefunctions of $\Psi_i(r), ~i=1,...,s$: \\
 \[  \Psi  = {\tt Chebyshev\_filter}(\Psi, ~m, ~b_{low}, ~b_{up}) ~,\]

\item Ortho-normalize the basis $\Psi$ by iterated Gram-Schmidt.
\item Perform the Rayleigh-Ritz (rotation) step: 
\begin{enumerate}[a)]
\item Compute $\hat H = \Psi^T H \Psi$; 
\item Compute the eigendecomposition of $\hat H$: ~$\hat H Q = Q D$, \\
      where  $D$ contains non-increasingly ordered eigenvalues of $\hat H$, \\
      and $Q$ contains the corresponding eigenvectors; 
\item 'Rotate' the basis as $\Psi := \Psi Q$; ~return $\Psi$ and $D$.
\end{enumerate}

\end{enumerate} 

\item  Goto step 3.

\end{enumerate} \vspace*{0.2cm}
}
\end{minipage} 
\tline
\end{algo}
\end{figure}

The upper bound at {\it step 7.1} in Algorithm \ref{chefsi} can be
obtained by using an upper-bound-estimator presented in \cite{chefsi}.
The {\tt Chebyshev-filter} step in {\it step 7.2} calls a subroutine
which applies the Chebyshev filter to each of the columns of $\Psi$.
If $m$ is the degree of the polynomial, this operation amounts to
computing the sequence of blocks $X_k$, $k=2, \cdots, m$
as follows: 
\[ 
X_{k+1} = {2 \over e} ( H -c I ) X_k - X_{k-1}, \quad k=1, 2,\cdots, m-1 
\]
starting with $X_0 = \Psi$, $X_1 = {1 \over e} (H - c I) X_0$. 
The returned filtered
block is $\Psi = X_m$. The scalars $e$ and $c$ are defined
by $e = (b_{up} - b_{low})/2$ and $c = (b_{up} + b_{low})/2$.
For simplicity we presented here an unscaled version of the filtering
process. 
To prevent the $X_k$ blocks from overflowing it  is safer
to scale them at each iteration. The scaling operation is inexpensive as it uses 
only values of the Chebyshev polynomial at the approximate smallest 
eigenvalue of the Hamiltonian. The reader is referred to 
\cite{chefsi} for details.  For discussion of scaling related to 
Chebyshev filtering, we refer interested readers to \cite{chebydav} or
a more detailed technical report
\cite{tr-pchefsi}.

The parallel implementation of Algorithms \ref{chefsi} 
is straightforward  with  the parallel  paradigm discussed  in
Section  \ref{ppp}. We  only mention  that the  matrix-vector products
related  to  filtering,  computing  upper  bounds,  and  Rayleigh-Ritz
refinement, can  easily be executed in  parallel. The re-orthogonalization
at {\it Step 7.3} ~of Algorithm \ref{chefsi} uses a parallel version of the
iterated Gram-Schmidt DGKS  method \cite{dgks:76}, which scales better
than the standard modified Gram-Schmidt algorithm.

The estimated  complexity of the algorithm  is similar to  that of the
sequential  CheFSI  method  in  \cite{chefsi}.  
For parallel computation it suffices
to estimate  the complexity  on a single  processor.  Assume  that $p$
processors are  used, i.e.,  each processor shares  $N/p$ rows  of the
full Hamiltonian.  The estimated  cost of a CheFS step on each
processor with respect to the  dimension of the Hamiltonian denoted by
$N$, and the number of computed states $s$, is as follows:
\begin{itemize}
\item The Chebyshev filtering in {\it Step 7.2} costs $O(s*N/p)$ 
flops. The discretized Hamiltonian is sparse and each matrix-vector 
product on one processor costs $O(N/p)$ flops. {\it Step 7.2} requires  
$m*s$ matrix-vector products, at a
total cost of $O(s*m*N/p)$ where the degree $m$ of the polynomial 
is small (typically between 8 and 20).

\item The ortho-normalization in {\it Step 7.3} costs $O(s^2*N/p)$
flops. There are additional communication costs 
because of the global reductions.

\item The eigen-decomposition at  {\it Step 7.4} costs $O(s^3)$ flops.

\item The final basis refinement step ($\Psi:= \Psi Q$) costs 
$O(s^2*N/p)$. 
\end{itemize}

If a  standard iterative diagonalization  method is used to  solve the
eigenproblem  \ref{kseq} at each  SCF step, then  it also
requires (i) the orthonormalization of a (typically larger) basis; (ii)
the eigen-decomposition of the projected Rayleigh-quotient matrix; and
(iii) the  basis  refinement (rotation).  
These  operations need  to  be  performed
several  times  within  this  single diagonalization.   But  
CheFS
performs each  of these operations only once  per SCF step.
Therefore, although  CheFS
scales  in a similar  way to
standard diagonalization-based  methods, the scaling  constant is much
smaller.  For large problems, CheFS  can achieve a tenfold or more 
speedup per SCF step,  over   using   the  well-known efficient
eigenvalue packages such as
ARPACK \cite{lesy:98} and TRLan \cite{trlan99,wusi:00}.
The total speedup can be more significant since 
self-consistency requires several SCF iteration steps. 

To summarize, a standard SCF method would have an outer SCF loop---the
usual  nonlinear SCF loop,  and an  inner diagonalization  loop, which
iterates   until   eigenvectors   are   within   specified   accuracy.
Algorithm~\ref{chefsi}  simplifies  this  by merging  the  inner-outer
loops into  a single  outer loop,  which can be  considered as  a {\it
nonlinear  subspace iteration  algorithm}.  The  inner diagonalization
loop is reduced into a single Chebyshev subspace filtering step.

\section{Chebyshev-Davidson algorithm for the first SCF iteration} 
 \label{chdav}
Within CheFSI, the  most expensive SCF step is  the first one, as it
involves a  diagonalization  in  order to  compute a  good
initial subspace to  be used for latter filtering.   In principle, any
effective eigenvalue  algorithms can  be used.  PARSEC  originally had
three  diagonalization  methods:  Diagla,  which is  a  preconditioned
Davidson method \cite{Saad-al-Elec,sosck:00};  the symmetric eigensolver 
in ARPACK  \cite{sorens:92,lesy:98};  and  
the Thick-Restart Lanczos algorithm called 
TRLan  \cite{trlan99,wusi:00}.
For systems of moderate sizes,  Diagla works well,  and then   becomes less
competitive relative to 
ARPACK or TRLan for larger systems when  a large number
of  eigenvalues are required.   
TRLan  is about twice as fast  as the symmetric
eigensolver  in  ARPACK, 
because  of  its reduced need for
re-orthogonalization.
In   \cite{chefsi},  TRLan  was used for
the diagonalization at the first SCF step.

For  very  large  systems,  memory  can become  a  severe  constraint.
One has to use eigenvalue algorithms with restart  since
out-of-core operations can be too slow.
However, even with standard restart  methods such as ARPACK and TRLan,
the  memory demand  can still  surpass  the capacity  of some
supercomputers. For example,  the $Si_{9041}H_{1860}$ cluster by TRLan
or ARPACK would require more memory than the largest memory
allowed for a job at the Minnesota  Supercomputing Institute in 2006.  
Hence it
is important to  develop a diagonalization method that  is less memory
demanding but whose efficiency is comparable to ARPACK and TRLan.  The
Chebyshev-Davidson method \cite{chebydav,bdavio} is developed with these
two goals in mind.

It  is generally  accepted that for  the implicit  filtering in  ARPACK and
TRLan to work efficiently, one  needs to use a subspace with dimension
about  twice  the  number  of  wanted eigenvalues.  This  leads  to  a
relatively large demand in memory when the number of
wanted eigenvalues  is large.  The  block Chebyshev-Davidson method
discussed  in \cite{bdavio} introduced an  {\it inner-outer  restart }
technique.   The {\it outer  restart} corresponds to  a standard
restart in which the subspace is truncated to a smaller dimension when
the specified  maximum subspace dimension is reached.   The {\it inner
restart}  corresponds to a  standard restart  restricted to  an active
subspace, it is performed when the active subspace dimension exceeds a
given  integer $act_{max}$ which  is much  smaller than  the specified
maximum  subspace  dimension.   With  {\it inner-outer  restart},  the
subspace  used  in Chebyshev-Davidson  is about half  the
dimension of the subspace required by ARPACK or TRLan.

We adapted  the proposed Chebyshev filters 
into a  Davidson-type eigenvalue  algorithm.  Although no  Ritz values
are available from previous SCF steps to be used as lower bounds, the
Rayleigh-Ritz refinement step within a Davidson-type method can easily
provide  a  suitable  lower  bound at each iteration.   
The upper  bound  is again estimated by
the upper-bound-estimator in \cite{chefsi},
and it is computed only once.   These two bounds are sufficient
for constructing  a filter at each  Chebyshev-Davidson iteration.  The
constructed filter magnifies the wanted  lower end of the spectrum and
dampens  the unwanted  higher  end, therefore  the  filtered block  of
vectors  have strong  components  in the  wanted eigensubspace,  which
results  in an  efficiency that  is comparable  to that  of  ARPACK or
TRLan.  The  main  structure  of  this  Chebyshev-Davidson  method  is
sketched  in  Algorithm  \ref{bcd},  we refer  interested  readers  to
\cite{bdavio} for algorithmic details.

\begin{figure}[htpb]
\begin{algo} Structure outline of the block Chebyshev-Davidson method 
  \label{bcd} \\
\tline
\vspace*{0.1cm}
\begin{minipage}[r]{\columnwidth}
{\baselineskip=4.6mm
\begin{enumerate}[~~1.~]

\item 
Compute ~${b_{up}}$ using the upper-bound-estimator in \cite{chefsi};~  
set ~${ b_{low}}$ as the median of the \\eigenvalues of the tri-diagonal matrix 
from the upper-bound-estimator.\\
Make the given initial size-$k$ block $V_1$ orthonormal, 
set $V = [V_1] $.

\item $[~ V_f ~]= { {\tt Chebyshev\_filter}} (V_1,  m, {b_{low},  b_{up}} )$.
\item Augment the basis $V$ by $V_f$:~ $V \leftarrow \left[ ~V, ~{V_f} ~\right]$, make $V$
orthonormal.

\item Inner-restart if active subspace dimension exceeds a given integer ~$act_{max}$.
\item Rayleigh-Ritz refinement:~ update matrix $M$ s.t. 
      ~$ M = V^T H V$;\\ 
      do eigendecomposition of $M$:~ $M~ Y = Y D $;~ 
      updated basis $V$:~ $V \leftarrow V \ Y$.

\item Compute residual vectors, determine convergence; \\
      perform deflation if some eigenpairs converge.

\item If all wanted eigenpairs converged, stop;  ~else, 
      adapt  ${b_{low}} = max(diag(D))$, \\
      set~ $V_1 =$ $[$ {\em the first $k$ non-converged Ritz vectors} in $V$ $]$.

\item Outer-restart if size of ~$V$ exceeds maximum subspace dimension.   
\item Continue from {step 2.}  
\end{enumerate}
\vspace*{0.2cm}
}
\end{minipage}
\tline
\end{algo}
\end{figure}

The first step diagonalization by the block Chebyshev-Davidson method,
together with the Chebyshev-filtered subspace (CheFS) method,
enabled us to perform SCF calculations for a class of large systems, 
including the silicon cluster $Si_{9041}H_{1860}$ for which 
over 19000 eigenvectors of a Hamiltonian with dimension around 
3 million were  to be computed.
These systems are practically infeasible with the 
other three eigensolvers (ARPACK, TRLan and Diagla) in PARSEC, using
the current supercomputer resources available to us at the 
Minnesota Supercomputing Institute (MSI).

\section{Numerical Results} \label{numer}

PARSEC has been applied to study a wide range of material systems 
(e.g. \cite{wjkc:04,parsec-k,ctws:94}).
The focus of this section is on large systems where relatively few
numerical results exist because of the infeasibility of 
eigenvector-based methods.
We mention that \cite{zhaoy-04} contains very 
interesting studies on clusters containing up to 
1100 silicon atoms, using the well-known efficient plane-wave
DFT package VASP \cite{kres-haf94,vasp96}; however, it is stated
in \cite{zhaoy-04} that a cluster with 1201 silicon atoms is 
``too computationally intensive''.
As a comparison, PARSEC using CheFSI, together
with the currently developed symmetric operations of real-space 
pseudopotential methods \cite{tiago_unpub},
 can now routinely solve
silicon clusters with several thousand atoms. 

The hardware  used for  the computations is  the SGI Altix  cluster at
MSI, it consists of 256 Intel Itanium processors at CPU rates of 1.6 GHz,
sharing 512  GB of memory (but a single job is allowed  to request at
most 250 GB memory).

The goal of the computations is not to study the parallel 
scalability of PARSEC, but rather to use PARSEC to do SCF calculation for 
large systems 
that were not studied before. Therefore we do not use different
processor numbers to solve the same problem. 
Scalability is studied in \cite{sosck:00} for the preconditioned 
Davidson method. Here we mention
that the scalability of CheFS is better than eigenvector-based
methods because of the reduced reorthogonalizations. 

\begin{table*}[hpt]  
\begin{center}
\begin{tabular}{|c|c|c|c|c|c|c|c|}  \hline  
sysem & dim. of $H$ &  $n_{state}$
& {\tt \#MVp} & {\tt \#SCF} & 
{\tt {total\_eV}/{atom}} & {\tt 1st CPU} & {\tt total CPU}  \\\hline
$Si_{2713}H_{828}$ $^a$
 & 1074080 & 5843
& 1400187 & 14 & -86.16790 &
 7.83 hrs. & 19.56 hrs. 
\\
$Si_{4001}H_{1012}$ $^b$ 
& 1472440 & 8511
& 1652243  & 12 & -89.12338 & 
18.63 hrs. & 38.17 hrs.
\\  
$Si_{6047}H_{1308}$ $^c$ 
& 2144432  & 12751
&  2682749  & 14  & -91.34809  & 
45.11 hrs. & 101.02 hrs.
\\  
$Si_{9041}H_{1860}$ $^d$ 
& 2992832 & 19015
& 4804488  & 18  &  -92.00412 & 
102.12 hrs. &  294.36 hrs
\\  
$Fe_{302}$ $^e$ 
& 2790688  & $1812 \times 2$  & 9377435 & 110  &  -795.18064 &
16.16 hrs. & 112.44 hrs.
\\ 
$Fe_{326}$ $^f$ 
& 2985992  & $1956 \times 2$  
& 10241385  & 119  &  -795.19898 &
11.62 hrs.  & 93.15 hrs.
\\ 
$Fe_{360}$ $^g$ 
& 3262312  & $2160 \times 2$  & 12989799   & 146  &  -795.22329 &
16.55 hrs.  & 140.68 hrs.
\\ 
\hline  
\end{tabular} 
{\footnotesize

\flushleft $^a$ $m=10$ for CheFS. 
First step diagonalization by TRLan cost 8.65 hours, projecting
it into a 14-steps SCF iteration cost around 121.1 hours.

\flushleft $^b$ First step diagonalization by TRLan cost 34.99 hours, projecting
it into a 12-steps SCF iteration cost around 419.88 hours.

\flushleft $^c$ Using 32 processors.

\flushleft $^d$ Using 48 processors.

\flushleft $^e$ $m=20$ for Chebyshev-Davidson;  $m=19$ for CheFS.

\flushleft $^f$ using 24 processors. 
$m=20$ for Chebyshev-Davidson;  $m=19$ for CheFS. 

\flushleft $^g$ using 24 processors. 
$m=20$ for Chebyshev-Davidson;  $m=17$ for CheFS.

}
\caption{Performance of the CheFSI method in various test systems. All
calculations were performed using 16 processors, and polynomial degrees $m=17$ for the
Chebyshev-Davidson and $m=8$ for CheFSI, except when otherwise stated.}
\label{si-fe} \end {center}
\end{table*}

In  the reported numerical  results, the  {\tt total\_eV/atom}  is the
total energy  per atom  in electron-volts, this  value can be  used to
assess accuracy of the final result; the {\tt \#SCF } is the iteration
steps needed to reach self-consistency; and the {\tt \#MVp} counts the
number of matrix-vector products.  Clearly {\tt \#MVp} is not the only
factor that  determines CPU time, the orthogonalization  cost can also
be a significant component.

For  all  of   the  reported  results  for  CheFSI,   the  first  step
diagonalization   used   the   Chebyshev-Davidson  method   (Algorithm
\ref{bcd}).   In  Table \ref{si-fe},  the  {\tt 1st  CPU}
denotes  the CPU  time  spent  on the  first  step diagonalization  by
Chebyshev-Davidson;  the {\tt  total CPU}  counts the  total  CPU time
spent to reach self-consistency by CheFSI.
 
\begin{table}[htp]
\begin{center}
\begin{tabular}{|c|c|c|c|c|}  \hline  
method &  {\tt \#MVp} & {\tt \#SCF} steps  & {\tt total\_eV/atom}  & CPU(secs) \\  \hline
CheFSI & 189755  &  11 &  -77.316873 & 542.43   \\  \hline
TRLan  & 149418  &  10 &  -77.316873 & 2755.49  \\  \hline
Diagla & 493612  &  10 &  -77.316873 & 8751.24  \\  
\hline  
\end{tabular}
\caption{ $Si_{525}H_{276}$, using 16 processors. 
The Hamiltonian dimension is 292584, 
where 1194 states need to be computed at each SCF step.
The first step diagonalization by Chebyshev-Davidson 
cost 79755 \#MVp and 221.05 CPU seconds; 
so the total \#MVp spent on CheFS in CheFSI is 110000. 
The polynomial degree used is $m=17$ 
for Chebyshev-Davidson and $m=8$ for CheFS.
The fist step diagonalization by TRLan requires 14909 \#MVp and 265.75
CPU seconds.}
\label{si525}
\end {center}
\end{table}

The first example in Table \ref{si-fe} is a relatively small silicon cluster 
$Si_{525}H_{276}$, which is used to compare the performance of CheFSI with
two eigenvector-based methods. 
All methods use the same symmetry operations \cite{tiago_unpub} in PARSEC.

For larger clusters $Si_{2713}H_{828}$  
and $Si_{4001}H_{1012}$ , Diagla became too slow to be practical.
However, we could still apply TRLan for the first step diagonalization 
for comparison, but we did not iterate until self-consistency was reached 
since that would cost a significant amount of our CPU quota. 
Note that with the problem size increasing, Chebyshev-Davidson compares
more favorably over TRLan. 
This is because we employed an additional trick in Chebyshev-Davidson, which
corresponds to allowing  
the last few eigenvectors not to converge
to the required accuracy.
The number of the non fully converged eigenvectors is bounded above by 
$act_{max}$, which is the maximum dimension of the active subspace. Typically 
$30 \le act_{max} \le 300 $ for Hamiltonian size over a million where several
thousand eigenvectors are to be computed. 
The implementation of this trick
is rather straightforward since it corresponds to applying the CheFS method
to the subspace spanned by the last few vectors in the basis 
that have not converged to required accuracy. 

For even larger clusters $Si_{6047}H_{1308}$  
and $Si_{9041}H_{1860}$, it became impractical 
to apply TRLan for the first step diagonalization 
because of too large memory requirements. For these large systems,
using an eigenvector-based method for each SCF step is clearly
not feasible. 
We note that the cost for the first step diagonalization by 
Chebyshev-Davidson is still rather high, it took close to 50\%
of the total CPU. 
In comparison, the CheFS method
saves a significant amount of CPU for SCF calculations over 
diagonalization-based methods, even if very efficient eigenvalue
algorithms are used.

\begin{figure}[htpb]
{\centering\epsfig{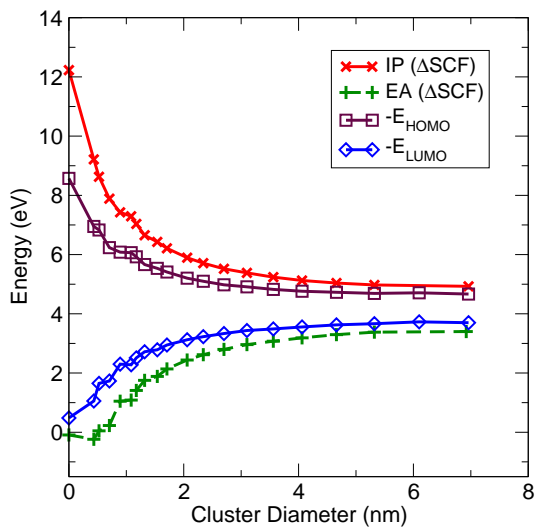}}
\caption{Ionization potential (IP, crosses) and electron affinity (EA,
  ``plus'' signs) for various clusters with diameters ranging from 0 nm
  ($SiH_{4}$) to 7 nm ($Si_{9041}H_{1860}$). Squares denote the
  negative of the highest occupied eigenvalue energy ($-E_{HOMO}$) of
  the neutral cluster. Diamonds denote the negative of the
  lowest unoccupied eigenvalue energy ($-E_{LUMO}$).} 
\label{si10k}
\end{figure}

Once the DFT problem, Eq. (\ref{kseq}), is solved, we have access to
several physical quantities. One of them is the ionization potential (IP)
of the nanocrystal, defined as the energy required to remove one
electron from the system. Numerically, we use a $\Delta SCF$ method:
perform two separate calculations, one for the neutral cluster and
another for the ionized one, and observe the variation in total energy
between these calculations. Figure \ref{si10k} shows the IP of several
clusters, ranging from the smallest possible ($SiH_{4}$) to
$Si_{9041}H_{1860}$. For comparison, we also show the eigenvalue of
the highest occupied Kohn-Sham orbital, $E_{HOMO}$. A known fact of
DFT-LDA is that the negative of the $E_{HOMO}$ energy is lower than the IP in
clusters \cite{martin_book_04}, which is confirmed in Figure
\ref{si10k}. In addition, the figure shows that the IP and $-E_{HOMO}$
approach each other in the limit of extremely large clusters.

Figure \ref{si10k} also shows the electron affinity (EA) of the
various clusters. The EA is defined as the energy released by the
system when one electron is added to it. Again, we calculate it by
performing SCF calculations for the neutral and the ionized systems
(negatively charged instead of positively charged now). In PARSEC,
this sequence of SCF calculations can be done very easily by reusing
previous information: The initial diagonalization in the second SCF
calculation is waived if we reuse eigenvectors and eigenvalues from a previous
calculation as initial guesses for the ChebFSI method. Figure \ref{si10k}
shows that, as the cluster grows in size, the EA approaches the
negative of the lowest-unoccupied eigenvalue energy. 
A power-law analysis in Figure \ref{si10k} indicates that both the
ionization potential and the electron affinity approach their bulk
values according to a power-law decay $R^n$ with exponent close
to 1. The numerical fits are:

\begin{equation}
{\rm IP} = {\rm IP}_0 + A/D^\alpha
\end{equation}

\begin{equation}
{\rm EA} = {\rm EA}_0 - B/D^\beta
\end{equation}
with ${\rm IP}_0 =$ 4.50 eV, ${\rm EA}_0 =$ 3.87 eV, $\alpha =$ 1.16,
$\beta =$ 1.09, $A =$ 3.21 eV, $B =$ 3.13 eV. These values for $A$ and
$B$ assume a cluster diameter $D$ given in nanometers. The difference
between ionization potential and electron affinity is the electronic
gap of the nanocrystal. As expected, the value of the gap extrapolated
to bulk, ${\rm IP}_0 - {\rm EA}_0 =$ 0.63 eV, is very close to the
energy gap predicted in various DFT calculations for silicon, which
range from 0.6 eV to 0.7 eV \cite{martin_book_04,aulbur00}. Owing to
the slow power-law decay, the gap at the largest crystal studied is
still 0.7 eV larger than the extrapolated value.

\begin{figure}[htpb]
{\centering\epsfig{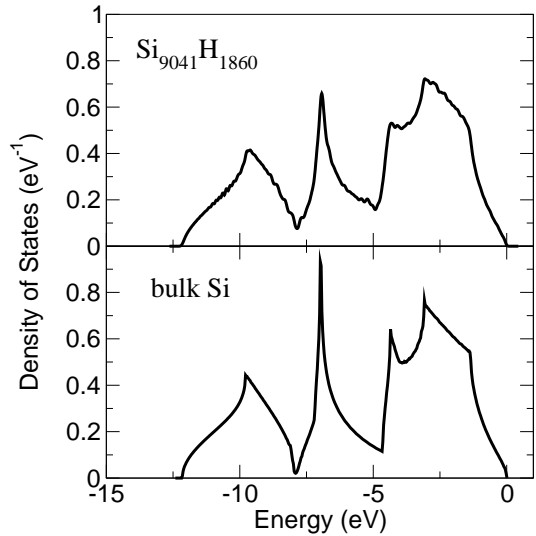}}
\caption{Density of states (DOS) of the
cluster $Si_{9041}H_{1860}$ (upper panel) compared with periodic
crystalline silicon (lower panel). 
As a consequence of the large size, the DOS of the
$Si_{9041}H_{1860}$ cluster is very close to that of bulk silicon 
(the infinite-size limit).} 
\label{si9k-dos}
\end{figure}

Other properties of large silicon clusters are also
expected to be similar to
the ones of bulk silicon, which is equivalent to a nanocrystal of
``infinite size''. Figure \ref{si9k-dos} shows that the density of
states already assumes a bulk-like profile in clusters with around ten
thousand atoms. The presence of hydrogen atoms on the surface is
responsible for subtle features in the DOS at around -8 eV and -3
eV. Because of the discreteness of eigenvalues in clusters, the DOS is
calculated by adding up normalized Gaussian distributions located at
each calculated energy eigenvalue. In Figure \ref{si9k-dos}, we used
Gaussian functions with dispersion of 0.05 eV.  More details are 
discussed in \cite{silicon-parsec}.

We also applied PARSEC to some large iron clusters. 
Extensive analysis of the magnetic properties of iron clusters based on
the methodology presented here and in previous work\cite{chefsi}, has
provided decisive evidence for surface effects in the magnetic moment of
these systems \cite{iron-parsec}, confirming earlier experimental data.
Table 
\ref{si-fe} also contains three clusters with more than 300 iron
atoms. 
These metallic systems are well-known to be very difficult for DFT
calculations, because of the ``charge sloshing'' \cite{payne:92,vasp96}.
The LDA approximation used to get exchange-correlation potential $V_{XC}$
is also known not to work well for iron atoms.
However, PARSEC was able to reach self-consistency for these large metallic 
clusters within reasonable time length. 
It took more than 100 SCF steps to reach self-consistency, 
which is generally considered too high for SCF calculations, but
we observed (from calculations performed on smaller iron clusters)
that eigenvector-based methods also required a similar number of
SCF steps to converge, thus the slow convergence is associated with
the difficulty of
DFT for metallic systems. Without CheFS, and under the same hardware conditions
as listed in Table \ref{si-fe},  
over 100 SCF steps using
eigenvector-based methods would have required months to complete
for each of these clusters. 

\section{Concluding Remarks} 

We developed and implemented the parallel CheFSI method for DFT SCF 
calculations. Within CheFSI, only the first SCF step requires a
true diagonalization, and we perform this step by the block 
Chebyshev-Davidson method. No diagonalization is required after the
first step; instead, Chebyshev filters are adaptively constructed to
filter the subspace from previous SCF steps so that the filtered subspace
progressively approximates the eigensubspace corresponding to occupied
states of the final Hamiltonian. 
The method can be viewed as a nonlinear subspace iteration method
which combines the SCF iteration and diagonalization, with the
diagonalization simplified into a single step Chebyshev subspace 
filtering.

Additional tests not reported here, have also shown that the subspace
filtering method is robust with respect to the initial
subspace. Besides self-consistency, it can be used together with
molecular dynamics or structural optimization, provided that atoms
move by a small amount. Even after atomic displacements of a fraction
of the Bohr radius, the CheFSI method was able to bring the initial
subspace to the subspace of self-consistent Kohn-Sham eigenvectors for
the current position of atoms, with no substantial increase in the
number of self-consistent cycles needed.

CheFSI significantly accelerates the  SCF calculations, and this enabled us
to perform  a class of large  DFT calculations that  were not feasible
before  by  eigenvector-based  methods.   As an  example  of  physical
applications, we discuss the  energetics of silicon clusters containing
up to several thousand atoms.

\acknowledgments

We thank the staff  members at the Minnesota Supercomputing Institute,
especially Gabe Turner, for  the technical support. There were several
occasions  where our large  jobs required  that the  technical support
staff change  certain default system  settings to suit our  needs. The
calculations  would  not  have  been  possible  without  the  computer
resource and the excellent technical support at MSI.
This work was supported by the MSI, by
the National Science Foundation under grants ITR-0551195, ITR-0428774,
DMR-013095 and DMR-0551195 and by the U.S. Department of Energy under
grants DE-FG02-03ER25585, DE-FG02-89ER45391, and
DE-FG02-03ER15491.



\end{document}